# Solar Flares and the Chromosphere

## *A white paper for the Decadal Survey**

L. Fletcher, R. Turkmani, H. S. Hudson, S. L. Hawley, A. Kowalski,
A. Berlicki, P. Heinzel

**Background:**
The solar chromosphere is at the center of many of the debates in solar physics. Highly complex, even in the quietest zones, the chromosphere is not an area that is easy to grasp. Across a few-Mm height the medium jumps from high plasma beta to low plasma beta, from neutral to ionized, from collisional to collisionless, becomes optically thin in the UV and mm-waves, and increases in temperature by almost three decades. It is no wonder that numerous instabilities occur. As observations improve, for example with the Hinode mission, these complexities become more apparent. Magnetic reconnection also occurs freely in this domain.

The chromosphere deserves attention in our attempts to find answers to the riddles of the corona, including flares. Understanding its magnetic complexity is a key to coronal modeling – particularly measurements of the vector field in the upper chromosphere are needed for a reliable reconstruction of coronal magnetic fields (e.g. DeRosa et al. 2009). The main provider of coronal heating may be in the transition region and upper chromosphere (Aschwanden et al 2007), and new evidence suggests that the chromosphere is the root of solar wind acceleration (De Pontieu et al. 2007).

> **Basic Characteristics**
>
> • The chromosphere and transition region anchor all coronal phenomena.
>
> • Current systems that stress the coronal field must penetrate the chromosphere.
>
> • A complex system of flows, shocks, jets, and waves originates in the chromosphere
>
> • Chromospheric radiation contains most of the flare luminosity

**The need for chromospheric observations of flares:**
The solar flare spectrum peaks in the optical and ultraviolet, making the solar chromosphere the dominant source of radiation in a solar flare (Canfield et al 1986, Neidig 1989, Woods et al. 2004). In other words, solar flares radiate most of their luminous energy in the chromosphere. The chromosphere is where electrons accelerated in flares lose their energy, and may also play a role in their acceleration (Fletcher & Hudson 2008, Brown et al. 2009). Yet despite its evident importance, observational knowledge of this region is relatively sparse. From spectroscopic observations of much more powerful stellar flares, in which the optical flare output dominates that of the whole star, we know the powerful diagnostic potential of chromospheric flare radiation, and existing observations already point to the utility of spatially-resolved optical/UV radiation in decoding the evolving flare magnetic field. There is a pressing need to return to the chromosphere to progress on understanding solar flares.

**Current Observations:**
Current ground-based efforts utilize small custom instruments on mid-sized telecopes and concentrate on (1) patrol images (Hα, He 10830A, white light) and (2) high resolution narrow-band spectroscopy of a few spectral lines (e.g. Hα, Ca 8542). There is little effort to conduct studies of chromospheric flares from space, though there has been notable success in targeted programs using TRACE, and serendipitous observations with Hinode, which have firmly established the relationship between the optical ('white light') footpoint, and the sites of strong non-thermal bremsstrahlung emission. It is to be expected that the Solar Dynamics Observatory will also reveal the detailed morphological evolution of ribbons, and possibly give some insight also into the structure of the bright ribbon sources.

**Information from imaging:**
It is well established that solar flare footpoints are very compact and evolve rapidly in position and intensity, reflecting in some detail the progression of flare energy release. Flare chromospheric ribbons are much more extended. Interpreted as the chromospheric intersections of evolving magnetic (quasi-) separatrix surfaces, flare ribbons allow the progression of field restructuring to be followed (Masson et al. 2009). Chromospheric flares have of course long been beautifully observed in H alpha, but also now in other lines such as the 304nm image from Hinode SDO (Figure 1). These show abundant fine detail from 3D structures. Flare ribbons give insight into both energetics and magnetic restructuring, in the impulsive and the gradual phase.

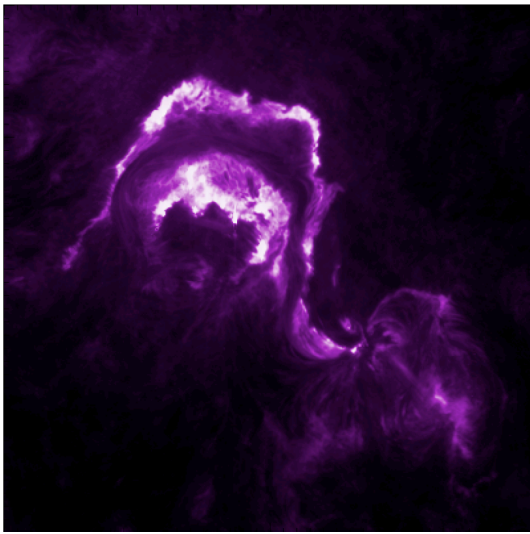

*Figure 1 – chromospheric flare ribbons observed with the Hinode AIA in the 304 nm channel (He II)*

Compared to chromospheric flare ribbons, the optical flare footpoints very little studied in the recent era. It is known however that a very precise timing and spatial relationship to the hard X-rays (Fig 2) and thus the flare electrons – the major energy-carrying component (Metcalf et al. 2003, Hudson et al 2006, Fletcher et al. 2007, Watanabe et al 2009). Relatively easily observed compared to hard X-rays, they provide a view of the total energy of a flare complementary to that provided by HXR radiation. White-light emission had previously been thought to be a 'big flare' phenomenon, but has now been seen even in small C-class events (Hudson et al 2006, Jess et al. 2009).

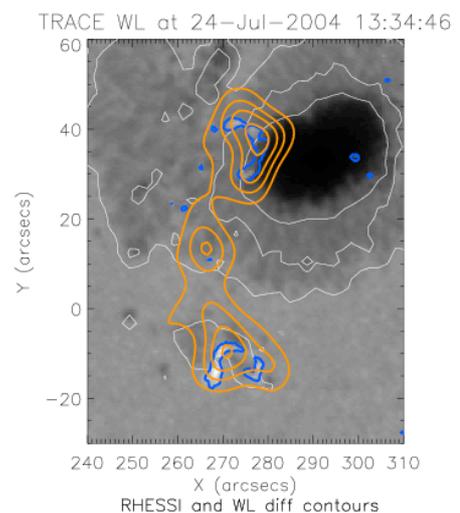

*Figure 2: RHESSI 25-50 keV contours of a flare (orange) and regions of strongest WL contrast (blue). There is an excellent spatial and temporal correlation, but the WL readily shows detail at the arcsecond level which remains difficult to obtain with HXRs. Image from Fletcher et al. 2007*

**Information from spectroscopy:**
During a flare, the chromosphere brightens in lines and continuum. At its most basic, continuum spectroscopic information is required for a proper assessment of the total radiative intensity in flares, which has so far been obtained only in a small number of events. Of particular interest is the identification (Neidig, 1989) of a Paschen jump (indicating the presence of the Paschen continuum from free-bound emission in the optical part of the spectrum) and flare-related changes to the Balmer and Paschen jumps in the flare spectrum, indicating variations in ionization fraction). Enhanced continuum shortwards of the Paschen jump may additional suggest increased Hα opacity, and photospheric excitation. This basic measurement has been made in very few flares (see e.g. Neidig 1983, Mauas 1990), and usually not for the strongest impulsive-phase optical flare kernels, but has major implications for the energy transport model (Canfield 1986, Neidig 1989, Fletcher et al 2007).

The chromospheric lines are potentially rich in diagnostic information. For example, Metcalf et al (1990a,b) have used Mg I lines to probe the temperature and density structure of the temperature minimum region. The Ca 8542 infrared triplet has been frequently used in stellar studies to diagnose the chromospheric temperature gradient, and the Balmer lines provide information on electron density (e.g. Svestka, 1976), and potentially also diagnostics of non-thermal electron distributions (Zharkova & Kobylinskii 1993, Kasparova & Heinzel 2002.) It has also been suggested that Lyα can also be used to detect flare proton fluxes in the 10-1000 keV energy range, with charge exchange on ambient neutrals resulting in shifted emission lines (Orrall & Zirker 1976). Never detected on the Sun, but known analogously in the terrestrial aurora, this would be a unique diagnostic of a major flare energy component as yet entirely invisible to us. The Balmer series line widths diagnose the electron density (via Stark broadening, e.g. Suemoto & Hiei 1959, Johns-Krull et al. 1997), and analysis implies lines imply that these flare lines are formed at electron densities in excess of $10^{13}$ cm$^{-3}$ (e.g. Johns-Krull et al 1997), meaning within a transient condensation. From spectroscopy we can also image plasma flows, extended to 3D with modeling.

**The poverty of solar flare chromospheric spectra:**
It is an embarrassment that there is still no imaging spectroscopy of solar flares from space, although the IRIS instrument will soon be available. There is also precious little in the way of ground-based spectra – regular measurements have not been made since the 1980s and early 90s. HMI on SDO is now providing imaging spectroscopy of a single photospheric line, Fe I 6173 A, and the EVE instrument provides full EUV spectra but of the Sun as a star. In the meanwhile, Figure 3 suggests the power of what may emerge from better spectroscopy: this is a stellar flare spectrum dominated by the Balmer series, but with white-light continuum (and photospheric lines) in the background (Kowalski et al. 2010). We do not have anywhere near comparable spectra for solar flares.

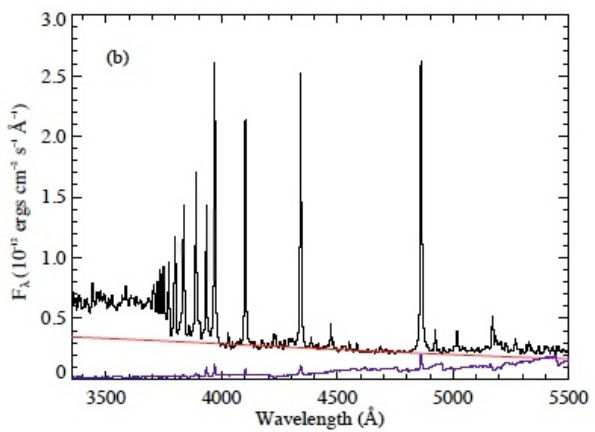

*Fig. 3 The blue spectrum of the flaring star YZ CMi (Kowalski et al. 2010). The red line indicates a hot blackbody component; most of the rest of the flare excess emission is the Balmer series plus the Balmer and Paschen continuum. The blue line is the quiescent star.*

**Implications for flare transport and acceleration:**
Line spectra give information about the physical conditions at the location where the line is excited, and can thus be used to understand how energy propagates through the chromosphere. At present, our understanding of solar flare particle acceleration rests on the collisional thick-target interpretation of a particle beam accelerated in the corona and propagating to the chromosphere where it loses energy monotonically with depth, due to Coulomb collisions. This makes firm predictions about the location of beam stopping, and chromospheric heating, ionization and line excitation. Systematic optical observations can confront this, and other, models. Line and continuum observations so far imply significant energy deposited in dense layers of the chromosphere, and while the total energy requirements can be met by the collisional thick target inferences (e.g. Watanabe et al 2010), in some cases the implication that the energy is concentrated at electron energies which cannot penetrate sufficiently deep in the chromosphere (Fletcher et al. 2007).

**Plasma physics:**
The physics of the chromosphere differs from that of the corona and photosphere because of its low level of ionization and the dominance of ion-neutral coupling at lower altitudes. The Earth's ionosphere has some points of analogy. The current systems that support coronal energy storage as non-potential structures must flow through the chromosphere, resulting in a much more complicated pattern of electrical conductivity than common modeling approaches (e.g. "line-tying") can comprehend. It would be a fair criticism that most flare modeling at present simply ignores this essential physics, as well as other factors likely to be important in chromospheric and transition-region dynamics.

**Seismic waves and Moreton waves:**
Large-scale wave structures originate in the lower solar atmosphere during flares. These include the global coronal waves responsible for radio type II bursts and Moreton waves. The seismic waves in the solar interior (Kosovichev and Zharkova 1998) now have been shown to be relatively commonplace (e.g. Donea & Lindsey 2005). Each of the global wave types offers powerful diagnostic information regarding the early development of the flare/CME process.

**New observations needed:**
Flare research in the fundamental domain of the chromosphere and transition region has only begun to advance beyond the classical H$\alpha$ stage. The text box summarizes the key observational goals for a new era in flare research. As Hinode and SDO demonstrate, space-based observations offer tremendous advantages in image stability, image dynamic range, and seeing. For the chromosphere and transition region, space also offers access to the key flare radiations at short wavelengths: spectra such as those of Figure 1 suggest temperatures above $10^4$ K for the recombination-radiation component, the dominant flare luminosity.

---

**Technical Requirements**

• Imaging at high spatial (0.1") and temporal (1 s) resolution in key optical and UV lines and continua, and with sufficient image dynamic range for flares.

• Ly$\alpha$ wing observations of charge-exchange signatures.

• Imaging spectroscopy of the chromosphere and transition region with full spectral coverage in optical-UV

• Modeling outside the "hydro-magnetic box": two-fluid, kinetic, 3D

We also recognize that ground-based facilities such as ATST will exist and produce key observations, and that ALMA (Karlicky et al. 2010) will bring an entirely new perspective of potentially great power.

**Summary:**